\documentclass[conference,letterpaper]{IEEEtran}

\usepackage[utf8]{inputenc} 
\usepackage[T1]{fontenc}
\usepackage{url, bm,amssymb}
\usepackage{ifthen}
\usepackage{graphicx}
\usepackage{cite}
\usepackage[cmex10]{amsmath} 

\begin{document}
\title{Distributed Stochastic Gradient Descent Using LDGM Codes} 



\author{%
  \IEEEauthorblockN{Shunsuke Horii\IEEEauthorrefmark{1},
  Takahiro Yoshida\IEEEauthorrefmark{2},
                     Manabu Kobayashi\IEEEauthorrefmark{3},
                     and Toshiyasu Matsushima\IEEEauthorrefmark{3}}
   \IEEEauthorblockA{\IEEEauthorrefmark{1}%
                     Waseda University,
                     Tokyo Japan,
                     s.horii@aoni.waseda.jp}
  \IEEEauthorblockA{\IEEEauthorrefmark{2}%
                     Yokohama College of Commerce, 
                     Kanagawa Japan,
                     t.yoshida@shodai.ac.jp}
   \IEEEauthorblockA{\IEEEauthorrefmark{3}%
                     Waseda University,
                     Tokyo Japan,
                     \{mkoba, toshimat\}@waseda.jp}
 }

\maketitle

\begin{abstract}
   We consider a distributed learning problem in which the computation is carried out on a system consisting of a master node and multiple worker nodes.
   In such systems, the existence of slow-running machines called stragglers will cause a significant decrease in performance.
   Recently, coding theoretic framework, which is named Gradient Coding (GC), for mitigating stragglers in distributed learning has been established by Tandon et al.
   Most studies on GC are aiming at recovering the gradient information completely assuming that the Gradient Descent (GD) algorithm is used as a learning algorithm.
   On the other hand, if the Stochastic Gradient Descent (SGD) algorithm is used, it is not necessary to completely recover the gradient information, and its unbiased estimator is sufficient for the learning.
   In this paper, we propose a distributed SGD scheme using Low Density Generator Matrix (LDGM) codes.
   In the proposed system, it may take longer time than existing GC methods to recover the gradient information completely, however, it enables the master node to obtain a high-quality unbiased estimator of the gradient at low computational cost and it leads to overall performance improvement.
\end{abstract}


\section{Introduction}

Recent advances in machine learning have achieved remarkable successes in various fields, such as image processing and natural language processing.
The amount of data processed by machine learning algorithms has been increasing dramatically, and it is difficult to process by a single computer or a single processor.
Therefore, the distributed computing system, in which data is distributed to many computers or processors and processed in parallel, is widely used.

Gradient-based methods such as Gradient Descent (GD) algorithm are one of the most widely used algorithms to fit the machine learning models over the training data. 
In order to handle massive amounts of data, developing distributed implementations of GD is important.
A common implementation of distributed GD is via a master/worker system where the data is distributed by a master node across multiple worker nodes.
Each worker computes a partial gradient based on its locally stored data and sends it to the master as soon as its computation is completed.
The master node aggregates all the partial gradients to update the model parameters.

In such systems, the master node needs to wait until all the worker nodes complete their computations and send their partial gradients.
Therefore, the run-time of each iteration of distributed GD is limited by slow-running workers, which is often called stragglers.

Recently, coding-theoretic strategies to mitigate stragglers have been attracting a lot of attention \cite{lee2018speeding,dutta2016short,yu2017polynomial,karakus2017straggler}.
They add some redundancy for the data to mitigate stragglers.
In particular, for the distributed GD, Gradient Coding (GC) has been proposed in \cite{tandon2017gradient}.
In the GC framework, the data is divided into some batches and the workers compute some partial gradients that correspond to the local data batches, and then send a linear combination of them.
By carefully designing the allocation of data batches and the linear combination coefficients, the master can recover the full gradient from a subset of workers' computation results.
There have been some further researches on GC to improve the performance \cite{li2018near,ye2018communication,halbawi2018improving}.

Most existing GC schemes are aiming at recovering the full gradient.
However, when the amount of data is tremendously large, an approximate gradient is often used.
Stochastic Gradient Descent (SGD) and its variants use an unbiased estimator of the full gradient \cite{shalev2014understanding,johnson2013accelerating,defazio2014saga}.
For SGD, the approximation accuracy of the approximate gradient determines the number of updates of the learning algorithm.
The authors in \cite{maity2018robust} proposed to use LDPC codes and iterative decoding algorithm in the GC framework.
They also indicated that the proposed scheme can be viewed as the SGD.
A disadvantage of their scheme is that it can only be applied to the case where the loss function is the squared loss.
The authors in \cite{raviv2017gradient} also proposed to use an approximate gradient in the GC framework.

In this paper, we propose a distributed SGD scheme using LDGM codes and peeling based decoding algorithm.
Although our work is similar to \cite{maity2018robust} in that a code with a sparse structure is used, the proposed scheme can be applied to loss functions other than the squared loss.
Another advantage of the proposed scheme is that the encoding and decoding complexity of it is very low.
In the proposed scheme, the obtained approximate gradient has a smaller approximation error compared to the case where no coding scheme is used and it results in the faster convergence of the learning algorithm.

The rest of the paper is organized as follows.
In Section 2, we introduce basic notations and definitions for the distributed learning problem.
In Section 3, we establish the distributed SGD scheme using LDGM codes and peeling based decoding algorithm.
A Density Evolution (DE) based analysis of the proposed scheme is also given.
In Section 4, we evaluate the effectiveness of the proposed scheme through numerical simulations.
Finally, we give a summary and future works in Section 5.

\section{Preliminaries}

In this section, we briefly review the model and definition of the distributed learning in a master/worker system.
Assume that we are given $n$ samples $\mathcal{D}=\left\{(\bm{x}_{i}, y_{i})\right\}_{i\in\left[n\right]}$, where $\bm{x}_{i}\in\mathbb{R}^{d}$ is a feature vector and $y_{i}\in\mathbb{R}$ is its label\footnote{In this paper, $[x]$ denotes $\left\{1,\ldots,x\right\}$}.
Let $\bm{w}\in\mathbb{R}^{d}$ be a parameter and $\ell(\bm{w}, \bm{x}, y)$ be a loss function for a sample $(\bm{x}, y)$\footnote{It is not necessary that the dimension of the parameter equals to that of the feature vector. However, for the sake of the simplicity, we assume that they are the same.}.
For example, if the linear model and squared loss is assumed, 
\begin{align}
\ell(\bm{w}, \bm{x}, y)=\frac{1}{2}(y-\bm{x}^{T}\bm{w})^{2}.\label{squared_loss}
\end{align}
We are interested in minimizing the following empirical loss function.
\begin{align}
\mathcal{L}(\bm{w})=\sum_{i=1}^{n}\ell(\bm{w},\bm{x}_{i}, y_{i})\label{empirical}
\end{align}
Since the empirical loss function is the sum of the loss function of each sample, the gradient of the empirical loss with respect to $\bm{w}$ has the following form.
\begin{align}
\nabla_{\bm{w}}\mathcal{L}(\bm{w})=\sum_{i=1}^{n}\nabla_{\bm{w}}\ell(\bm{w},\bm{x}_{i}, y_{i})
\end{align}
A popular approach to minimizing the empirical loss is via the GD.
The GD iteratively updates the estimated parameter vector $\bm{w}^{(t)}$ by moving along the negative gradient direction as follows.
\begin{align}
\bm{w}^{(t+1)}=\bm{w}^{(t)}-\eta^{(t)}\nabla_{\bm{w}}\mathcal{L}(\bm{w}^{(t)}),
\end{align}
where, $\eta^{(t)}$ is the learning rate in the $t$th iteration.

When the size of the training data is too large to process on a single machine or a single processor, one way to implement the GD updates is to calculate the gradient in a distributed fashion over many computing nodes.
We consider a master/worker system that consists of a master node and $N$ worker nodes.

Without any coding scheme, a naive implementation of the distributed GD is that we first divide the data into $N$ chunks $\left\{\mathcal{D}_{1},\ldots,\mathcal{D}_{N}\right\}$ of size $\frac{n}{N}$ and each chunk $\mathcal{D}_{j}$ is stored on worker $j$.
Within each iteration of the GD updates, the master broadcasts the current estimate $\bm{w}^{(t)}$ to all the workers and then each worker $j$ calculates $\sum_{(\bm{x},y)\in \mathcal{D}_{j}}\nabla_{\bm{w}}\ell(\bm{w}^{(t)}, \bm{x}, y)$, and sends it to the master.
The master waits for the results from all the workers and sums them up to obtain the full gradient
\begin{align}
\nabla_{\bm{w}}\mathcal{L}(\bm{w}^{(t)})=\sum_{j=1}^{N}\sum_{(\bm{x},y)\in \mathcal{D}_{j}}\nabla_{\bm{w}}\ell(\bm{w}^{(t)}, \bm{x}, y).
\end{align}
In this scheme, the master has to wait until all the workers complete their computations.
Therefore, even a single straggler can significantly delay the computation time in each iteration.

The GC scheme enables the system that the master can recover the full gradient with the results from a subset of workers by adding some redundancy on the data stored in the workers.
Here, we describe the GC scheme using an $(N, K)$ code.
The data divided into $K$ chunks $\left\{\mathcal{D}_{1},\ldots,\mathcal{D}_{K}\right\}$ of size $\frac{N}{K}$.
Let $\mathcal{N}(j)\subseteq \left[K\right]$ and assume that each worker $j$ stores $\left\{\mathcal{D}_{k}\right\}_{k\in\mathcal{N}(j)}$.
Then, each worker $j$ computes the following partial gradients
\begin{align}
\bm{g}_{k}=\sum_{(\bm{x},y)\in\mathcal{D}_{k}}\nabla_{\bm{w}}\ell(\bm{w}, \bm{x}, y),\quad k\in\mathcal{N}(j)
\end{align}
and sends their linear combination $\sum_{k\in\mathcal{N}(j)}b_{j,k}\bm{g}_{k}$ to the master.
Here, for the sake of simplicity, we drop the superscript $(t)$ denoting the iteration number of GD.
By carefully designing $\left\{\mathcal{N}(j)\right\}_{j\in\left[N\right]}$ and $\left\{b_{j,k}\right\}_{j\in\left[N\right], k\in\mathcal{N}(j)}$, the master can recover the full gradient based on the results from some fastest workers.
See \cite{tandon2017gradient} for more details.

\section{Distributed Stochastic Gradient Descent using LDGM codes}

\subsection{Encoding}

\begin{figure}[t]
   \centering
   \includegraphics[width=0.9\linewidth]{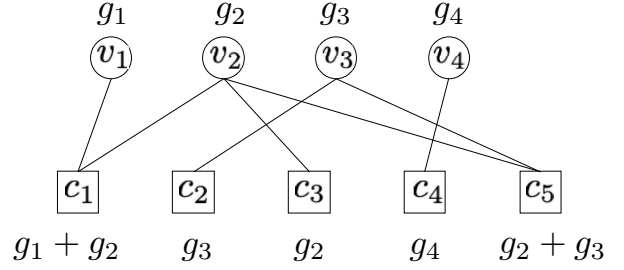}
   \caption{Tanner graph representation of the gradient coding scheme.}
   \label{fig:gc_graph}
\end{figure}

First, we present a bipartite graph representation of the GC scheme.
We consider a graph that consists of two sets of nodes $(\mathcal{V}, \mathcal{C})$, where $\mathcal{V}=\left\{v_{1},\ldots,v_{K}\right\}$ denotes the set of the partial gradients $\left\{\bm{g}_{k}\right\}_{k\in\left[K\right]}$ and $\mathcal{C}=\left\{c_{1},\ldots,c_{N}\right\}$ denotes the set of linear combinations of the partial gradients computed by workers.
In our scheme, the linear combinations are simply the sum of the partial gradients.
An edge is connected between $v_{k}$ and $c_{j}$ if $k\in\mathcal{N}(j)$.
An example of the graph is shown in Fig. \ref{fig:gc_graph} for $K=4$ and $N=5$.
In Fig. \ref{fig:gc_graph}, the circle nodes represent the partial gradients that need to be recovered and the square nodes denote the generator nodes which represent that the worker $j$ computes the sum of the partial gradients $\sum_{k\in\mathcal{N}(j)}\bm{g}_{k}$.
The graph can be seen as a Tanner graph for a low-density generator matrix (LDGM) code \cite{richardson2008modern}.
Here, the sum operation at the generator nodes is over the real filed vector, whereas in an LDGM code, the sum is over the finite field scalar.

An ensemble of LDGM codes is determined by degree distributions \cite{richardson2008modern}.
Let $L(x)=\sum_{i}L_{i}x^{i}$ and $\lambda(x)=\sum_{i}\lambda_{i}x^{i-1}$ denote the variable-node degree-distributions from the node and edge perspectives, respectively.
A variable node is connected with $i$ generator nodes with the probability $L_{i}$ and $\lambda(x)=\frac{L'(x)}{L'(1)}$.
Similarly, let $R(x)=\sum_{i}R_{i}x^{i}$ and $\rho(x)=\sum_{i}\rho_{i}x^{i-1}$ denote the generator-node degree-distributions from the node and edge perspectives, respectively.
A generator node is connected with $i$ variable nodes with the probability $R_{i}$ and $\rho(x)=\frac{R'(x)}{R'(1)}$.

\subsection{Decoding}

In our scheme, the master tries to recover a subset or all of the partial gradients $\left\{\bm{g}_{k}\right\}_{k\in\left[K\right]}$ by an iterative algorithm, which is similar to the peeling decoding algorithm for the binary erasure channel (BEC) using the Tanner graph.
For ease of analysis, we assume that workers that could not complete their computations within time $t_{0}$ as stragglers and the master starts running the decoding algorithm based on the computation results of other than stragglers. (In practice, the master can start running the decoding algorithm as soon as it receives computation results of the workers who have completed their computations.)
Each variable (generator) node sends an outgoing message along each edge connected to the generator (variable) node whose value is an erasure or a real value vector.
At a generator node of degree 1, if the corresponding worker is not a straggler, the outgoing message along the edge is the computation result itself that the corresponding worker computed.
At a generator node of larger degree, the outgoing message along the edge is not an erasure if the corresponding worker is not a straggler and the incoming messages along the other edges connected to that generator node are not erasures.
In this case, the outgoing message is the computation result of the corresponding worker minus the sum of the incoming messages along the other edges.
In cases other than the above cases, the outgoing message from a generator node is an erasure.
At a variable node, the outgoing message is an erasure if the incoming messages along all the other edges connected to the variable node are erasures.
Otherwise, the outgoing message is any one of the non-erasure incoming messages along the other edges.

\subsection{Density Evolution}

Let $\mathcal{G}(K, N, \lambda, \rho)$ denote the ensemble of Tanner graphs corresponding to the GC scheme with $K$ variable nodes, $N$ generator nodes, and the degree distribution pair $(\lambda(x), \rho(x))$.
We consider the decoding performance averaged over the ensemble of graphs $\mathcal{G}(K, N, \lambda, \rho)$ in the limit as $K, N\to \infty$.
To do so, we need some assumptions on the computation time of the workers.
We assume that the computation time $T_{j}$ of the worker $j$ is a random variable whose cumulative distribution function is $F_{j}(t)$ and it is independent to the computation times of other workers.
Further, we assume that the distribution function $F_{j}(t)$ satisfies $F_{j}(t)=F(t/|\mathcal{N}(j)|)$ for a base distribution function $F(t)$.
For example, if the computation times are modeled by exponential distribution, $F(t)=1-e^{-\mu t}$, where $\mu$ is a parameter that determines how long time is required for workers to complete their computations.
Above assumption reflects the fact that the more partial gradients have to be computed by a worker, the more computation time is required to complete the computation.
A similar assumption is made in \cite{lee2018speeding}.

Let $\texttt{x}_{l}$ and $\texttt{y}_{l}$ be the probabilities that an outgoing message from a variable node and a generator node, respectively, are erased during the $l$th iteration.
The depth-2$l$ neighborhood of a randomly chosen edge in $\mathcal{G}(K, N, \lambda, \rho)$ is tree-like with probability one as $K, N\to \infty$.
By considering the decoding algorithm, we obtain the following density evolution (DE) formula.
\begin{align}
\texttt{y}_{1} &= 1-\tilde{\rho}(0)\\
\texttt{x}_{l} &= \lambda(\texttt{y}_{l}),\quad l\ge 1\\
\texttt{y}_{l+1} &= 1-\tilde{\rho}(1-\texttt{x}_{l}),\quad l\ge 1
\end{align}
where $\tilde{\rho}(x)=\sum_{i}\tilde{\rho}_{i}x^{i-1}$ and
\begin{align}
\tilde{\rho}_{i}=\rho_{i}\left(1-F(t_{0}/i)\right).
\end{align}

\subsection{Stochastic Gradient Descent}
In the proposed scheme, by increasing the value of $t_{0}$, the master can recover the full gradient.
However, it may take very long time.
If we use SGD instead of GD, the master does not have to recover the full gradient.
The (mini-batch) SGD iteratively updates the parameter vector $\bm{w}$ as follows.
\begin{align}
\bm{w}^{(t+1)}=\bm{w}^{(t)}-\eta^{(t)}\sum_{i\in\mathcal{I}}\nabla_{\bm{w}}\ell(\bm{w}^{(t)}, \bm{x}_{i}, y_{i}),
\end{align}
where $\mathcal{I}\subseteq \left[n\right]$ and SGD is equivalent to GD if $\mathcal{I}=\left[n\right]$.
The approximate gradient term (second term of the right-hand side) can be interpreted as an unbiased estimator of the full gradient assuming a uniform distribution on the training data.
Note that the size of $\mathcal{I}$ trades the approximation error of the approximate gradient to the computational complexity to calculate it.

In our proposed scheme, even if the master fails to recover the full gradient, it could recover a subset of $\left\{\bm{g}_{k}\right\}_{k\in\left[K\right]}$.
Let $\mathcal{K}\subseteq \left[K\right]$ be the set of the partial gradients that the master obtains by the decoding algorithm, the master updates $\bm{w}$ as follows.
\begin{align}
\bm{w}^{(t+1)}=\bm{w}^{(t)}-\eta^{(t)}\sum_{k\in\mathcal{K}}\bm{g}_{k}
\end{align}

Note that if we use SGD for the learning algorithm, we can take the strategy that we use no GC scheme and ignore the computation results of stragglers.
However, by using GC scheme, we can expect that the approximation error of the approximate gradient is reduced.

\section{Experiments}

In this section, we present some experimental results of the proposed scheme.
In particular, we empirically compare the performance of our proposed distributed SGD using LDGM codes with the distributed GD using GC scheme in \cite{halbawi2018improving} and distributed SGD with the naive uncoded scheme where no redundancy among the workers is added.

\begin{figure*}[t]
\begin{center}
  \includegraphics[width=\linewidth]{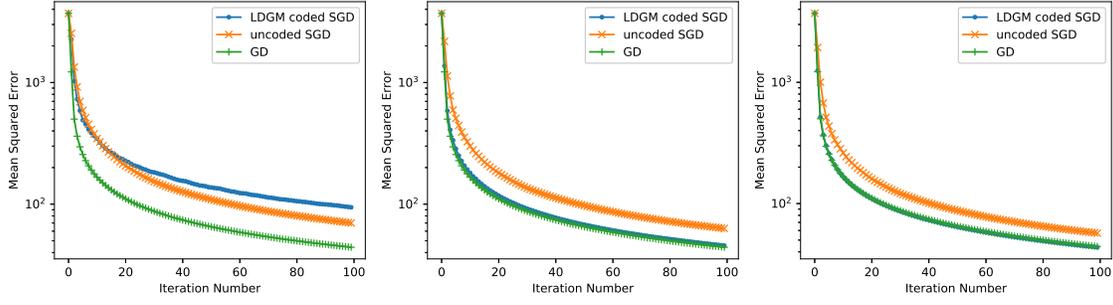}
  \caption{The values of the objective functions (mean squared error) as the functions of the iteration number for $\mu=0.5, 1.0$ and $2.0$.}
  \label{fig:iteration}
 \end{center}
\end{figure*}

\begin{figure*}[t]
\begin{center}
  \includegraphics[width=\linewidth]{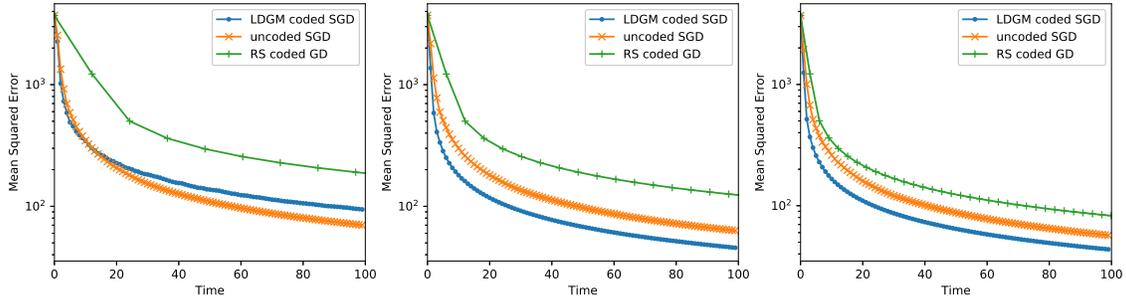}
  \caption{The values of the objective functions (mean squared error) as the functions of the processing time for $\mu=0.5, 1.0$ and $2.0$. Time of the master's decoding is not included.}
  \label{fig:time}
 \end{center}
\end{figure*}

\subsection{Experimental Setup}
We consider to solve a least square problem, that is, squared loss function (\ref{squared_loss}) is assumed.
The elements of each feature vector $\bm{x}_{i}$ are drawn from the standard normal distribution, and label $y_{i}$ is constructed by $y_{i}=\bm{x}_{i}^{T}\bm{w}^{*}+\epsilon$, where the elements of $\bm{w}^{*}$ and $\epsilon$ are also drawn from the standard normal distribution.
The dimension $d$ of the feature vectors and the sample size $n$ are set to $12000$ and $1200$, respectively.

We simulate the master/worker system with stragglers as follows.
As described in the part of DE, we assume that the computation time of each worker is a random variable.
In GC schemes, the computation time of worker $j$ follows $F(t/|\mathcal{N}(j)|)$, where $|\mathcal{N}(j)|$ is the number of chunks that the worker $j$ process and $F(t)$ is a base distribution function.
In this experiment, we assume that $F(t)=1-e^{-\mu t}$.
In our proposed scheme, we set the time threshold $t_{0}=1$ and a worker whose computation time is larger than $t_{0}$ is treated as a straggler.
Therefore, the probability that the worker $j$ is a straggler is $e^{-1/|\mathcal{N}(j)|}$.
In GC schemes, the data is divided into $K$ chunks and each worker computes the partial gradients of some chunks.
On the other hand, in uncoded system, the data is divided into $N$ chunks and each worker computes the partial gradient of a single chunk.
Therefore, the size of each chunk in uncoded system is different from that in GC schemes.
We assume that the computation time of each worker in uncoded system follows $F(Nt/K)$, where $N/K$ is the ratio of the batch size of uncoded system and that of GC schemes.

In our proposed scheme, we have to determine the degree distribution pair $(\lambda(x), \rho(x))$.
We searched the variable regular generator (check) irregular distribution pair based on DE so that (numerically) converged value of $\texttt{x}_{l}$ is minimized subject to the constraint that the rate $K/N$ of the code is $1/2$.
The found degree distribution pair is
\begin{align}
\lambda(x)=x^{2},\quad \rho(x)=\frac{3}{4}+\frac{1}{4}x^{2},
\end{align}
for $\mu=0.5$ and
\begin{align}
    \lambda(x)=x^{2},\quad \rho(x)=\frac{1}{2}+\frac{1}{2}x,
\end{align}
for $\mu=1.0, 2.0$.
In an environment with many stragglers, we found that LDGM codes with more degree 1 generator nodes are preferred.
We run the decoding algorithm until the algorithm converges.
Finally, we set the learning rate $\eta^{(t)}=\eta^{(0)}/t$ with $\eta^{(0)}=0.1$ for all schemes.

Fig. \ref{fig:iteration} shows how the value of the objective function (\ref{empirical}) of each method decreases with the number of iterations.
When $\mu$ is small, uncoded SGD scheme is better than LDGM coded SGD scheme.
This is because in a situation where workers' processing time is long and there are many stragglers, the decoding algorithm can not correct the erasures well.
In such a situation, it is more efficient to divide the data into many batches and reduce the batch size instead of coding\footnote{When coding is not performed, the size of each batch is $n/N$, while it is $n/K$ when coding is performed and $N>K$.}.
On the other hand, LDGM coded SGD scheme has a similar performance of (full) GD scheme when the probability of each worker is straggler is rather small, and the performance of it is much better than that of uncoded SGD scheme.
When the probability of each worker is straggler is small, LDGM coded scheme can recover almost all the partial gradients and in such a situation, the gain obtained by coding exceeds that obtained by reducing the batch size.

In a master/worker system with stragglers, we need a GC scheme in order to implement GD.
For comparison, we used GC scheme proposed in \cite{halbawi2018improving} (RS coded GD scheme).
The GC scheme in \cite{halbawi2018improving} has a parameter $w$, that is the number of batches that each worker processes.
The expectation of the time $T_{wait}$ of the master has to wait in each iteration depends on this parameter.
In our experiment setting, the expected wait time $E[T_{wait}]$ is expressed as 
\begin{align}
    E[T_{wait}]=\frac{\mu}{w}\left(1+\frac{1}{2}+\ldots+\frac{1}{N-\lfloor wN/K\rfloor+1}\right).
\end{align}
We searched $w$ that minimizes $E[T_{master}]$ and the optimal $w$ is $1$ for $\mu=0.5, 1.0, 2.0$ and $E[T_{wait}]$ is $12.112, 6.056, 3.028$, respectively.
Fig. \ref{fig:time} shows how the value of the objective function of each method decreases with processing time.
In this experiment, the decoding time of the master is not included because it depends on the implementation.
Therefore, the processing time in each iteration is $1.0$ for uncoded SGD scheme and LDGM coded SGD scheme .because we set $t_{0}=1.0$
We assume that the processing time in each iteration is $E[T_{wait}]$ for RS coded GD scheme.
In our experiment setup, the SGD schemes are better than the GD scheme with GC and the LDGM coded SGD scheme shows the best performance for $\mu=1.0, 2.0$.

\section{Conclusion}
We have developed a gradient coding scheme based on LDGM codes and iterative decoding algorithm.
We also developed a density evolution analysis of the proposed scheme.
Although the proposed system may require more time than existing gradient coding schemes to obtain the full gradient, it can recover an approximate gradient with high accuracy in a low computational complexity.
Combining the proposed scheme and the stochastic gradient descent (SGD) algorithm, we can obtain a distributed learning algorithm which converges faster than the full gradient descent with a gradient coding scheme.

There are some future directions of the work presented here.
In our experiment, we fixed the threshold parameter that the master waits for the workers' responses.
This parameter trades the approximation error of the approximated gradient with the master's waiting time.
We need a method to decide what value this parameter should be set in order to accelerate the convergence of the whole learning algorithm.
We run the decoding algorithm until it converges.
The number of the iteration of the decoding algorithm trades the approximation error of the approximate gradient with the master's decoding time.
We also need a method to determine the number of iterations of the decoding algorithm to accelerate the learning algorithm.

In our proposed scheme, we used the SGD algorithm for the learning algorithm.
There are some variants of the SGD such as SVRG and SAGA \cite{johnson2013accelerating, defazio2014saga}.
In these methods, the convergence of the learning algorithm is accelerated at the cost of computing an accurate gradient per an update of the parameter.
It is a future work to construct a high-performance distributed learning scheme by combining these learning algorithms and the proposed gradient coding scheme.

\section*{Acknowledgment}

We would like to acknowledge all members of Matsushima Lab. and Goto Lab. in Waseda Univ. for their helpful suggestions to this work.
This research is partially supported by No. 16K00417 of Grant-in-Aid for Scientific Research Category (C) and No. 18H03642 of Grant-in-Aid for Scientific Research Category (A), Japan Society for the Promotion of Science.


\bibliographystyle{IEEEtran}
\bibliography{ref}

\end{document}